\documentclass[aip,amsmath,amssymb,reprint,apl]{revtex4-1}

\usepackage{graphicx}
\usepackage{dcolumn}
\usepackage{bm}

\usepackage[utf8]{inputenc}
\usepackage[T1]{fontenc}
\usepackage{mathptmx}
\usepackage{color,soul}

\begin{document}

\preprint{AIP/123-QED}

\title{Pinhole quantum ghost imaging}

\author{Andres Vega}%
 \email{andres.vega@uni-jena.de}
 \affiliation{
Institute of Applied Physics, Abbe Center of Photonics, Friedrich Schiller University Jena, Albert-Einstein-Str. 15, 07745 Jena, Germany}%
\author{Sina Saravi}
\affiliation{
Institute of Applied Physics, Abbe Center of Photonics, Friedrich Schiller University Jena, Albert-Einstein-Str. 15, 07745 Jena, Germany}%
\author{Thomas Pertsch}
\affiliation{
Institute of Applied Physics, Abbe Center of Photonics, Friedrich Schiller University Jena, Albert-Einstein-Str. 15, 07745 Jena, Germany}%
\affiliation{
Fraunhofer Institute for Applied Optics and Precision Engineering IOF, Albert-Einstein-Str. 7, 07745 Jena, Germany}
\author{Frank Setzpfandt}
\affiliation{
Institute of Applied Physics, Abbe Center of Photonics, Friedrich Schiller University Jena, Albert-Einstein-Str. 15, 07745 Jena, Germany}%

\date{\today}

\newcommand{\ii}{{\rm i}}
\renewcommand{\d}{{\rm d}}
\newcommand{\x}{{\rm x}}
\newcommand{\y}{{\rm y}}
\newcommand{\z}{{\rm z}}
\renewcommand{\P}{{\rm P}}
\renewcommand{\S}{{\rm S}}
\newcommand{\I}{{\rm I}}

\begin{abstract}
We propose a quantum ghost imaging scheme based on biphotons, that, by using a collimated pump beam of the right size for biphoton generation, obviates the need for lenses to achieve imaging. The scheme is found to be analogous to the classical pinhole camera, where we show that the equivalent to the classical pinhole size depends mainly on the width of the pump beam, but also on the thickness of the nonlinear crystal and the wavelengths of the biphoton.
\end{abstract}

\maketitle

Quantum ghost diffraction \cite{shih1} and imaging \cite{shih2} rely on the spatial correlations of a biphoton, which can be created by the nonlinear process of spontaneous parametric down conversion (SPDC)\cite{SPDCExp, SPDCHongMandel}, where a pump ($\P$) photon impinging on a crystal with second-order nonlinearity $\chi^{(2)}$ is split into a pair of photons called signal ($\S$) and idler ($\I$). After creation, the two photons are separated into two different paths and only one of them interacts with the object, e.g. the signal photon. Then, the signal photon is measured by a detector with no spatial resolution, whereas another detector with spatial resolution measures the idler photon that never interacted with the object. None of the detectors alone can recover a diffraction pattern or image of the object. Remarkably, these can be retrieved by correlating the two measurements\cite{CoincidenceImaging1,CoincImaging2}.
This measurement technique has two main advantages. First, very low numbers of photons can be used 
due to the inherently better signal-to-noise ratio of quantum ghost imaging compared to imaging with classical light \cite{subshot,small}. Additionally, ghost imaging with two-color biphotons can overcome limitations due to inaccessible wavelength ranges for illumination and detection \cite{TwoColorBoyd,TwoColorShih,PhotonSparse}.

To form a ghost image, usually lenses are placed in the path of the signal and/or idler after the crystal \cite{shih2,small,subshot} or in the pump beam before the crystal \cite{shih_concavemirror}. The lenses introduce a parabolic phase-front in either of the beam paths, which results in the formation of the image in the coincidence measurement. However, quantum ghost imaging can be also realized without lenses by adding the parabolic phase-front through engineering the nonlinearity profile of the nonlinear crystal, e.g. by using a nonlinear photonic crystal \cite{lenslessphotonic}.
Furthermore, as ghost imaging can be also realized with classical light using thermal light sources\cite{RoleEntangQuantClass,GhostImagClass4,ClassConjMirror,GhostImagClass3,GhostImagClass2,GhostImagClassic1}, their inherent property of acting like a phase-conjugated mirror \cite{ClassConjMirror} in a ghost imaging scheme can also be used for lensless ghost imaging\cite{thermallensless,LenslessClassEqualArms}.

In classical optics, lensless imaging can be also realized using the principle of pinhole imaging \cite{pinhole,young1989}.
In a pinhole camera, the object is located on one side of an opaque screen with a small pinhole, whereas the detector is on the other side. Without the need of lenses, the detector captures a shadow of the object, which can be optimized by adapting the pinhole size\cite{pinhole}. Throughout this manuscript, we will refer to this shadow as an image although strictly no imaging is taking place.
For applications where high spatial resolution is not needed, this type of lensless imaging has several advantages over imaging with lenses, among which are a larger depth of field, a wide angular field of view\cite{young1989}, and its applicability in wavelength ranges for which high-quality lenses are less available\cite{XrayPinhole,GammaPinhole}.

\begin{figure}[b]
\centering
\includegraphics[scale=1]{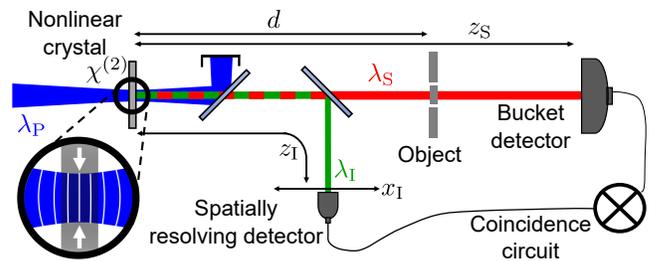}
\caption{\label{fig:setup} Sketch of the considered setup.}
\end{figure}

In this work, we want to show that the advantages of pinhole imaging can be also harnessed in quantum ghost imaging. For ghost imaging with thermal light, a pinhole-based scheme has been already proposed, where the optimal lensless imaging condition depends on the size of the thermal source \cite{thermalpinhole}.
We extend this approach of lensless imaging to the quantum regime with entangled photons based on the setup sketched in Fig.~\ref{fig:setup}. Here, we assume that the nonlinear crystal generating photon pairs is illuminated by a collimated pump beam and, contrary to ghost imaging with thermal light, we use a bucket detector instead of a point detector behind the object. We will show, that for specific pump beam diameters, pinhole quantum ghost imaging can be realized and we investigate its properties and optimum regime of operation. 
To this end, we start by discussing the biphoton joint spatial probability (JSP) and the quantum ghost pattern ($G$) together with a numerical example. Later, we derive a simplified analytical model for our imaging scheme that explains the observations of the numerical example and allows the connection to the classical pinhole camera. Using this model, we will furthermore discuss the spatial resolution of pinhole quantum ghost imaging.

Throughout this work, $z$ is the propagation direction and we restrict our analysis, without loss of generality, to one transverse dimension $x$ in position space, whose conjugate in momentum space is the transverse component of the wave-vector $k_x$. We also assume an infinitely extended nonlinear crystal in the transverse direction, which ensures transverse phase matching $k_{x\P} = k_{x\S}+k_{x\I}$.
We assume an undepleted monochromatic classical pump beam of the form $E_P (x,z,t) = \int \d k_{x\P} \:  \phi_\P (k_{x\P}) \: \exp{[i (k_{x\P} x + k_{z\P} z - \omega_\P t)]} $, where  $\phi_\P (k_{x\P})$ is the pump spatial spectrum, the longitudinal component of the wave-vector is $k_z = [(\omega n/c)^2 - k_x^2]^{1/2}$ with $\omega/c = 2\pi/\lambda$, $\lambda$ being the wavelength in free space and $\omega$ the corresponding frequency. Additionally, we ignore the effects of the boundaries between the crystal and its surrounding free space, which would cause refracted and reflected waves, and assume a simplified case with the crystal's refractive index $n=1$.
We investigate signal and idler photons at fixed frequencies of $\omega_\I$ and $\omega_\S$, such that $\omega_\P = \omega_\S + \omega_\I$. This can be achieved experimentally by placing narrow bandpass filters centered around these frequencies in their beam paths.
Under these conditions, the biphoton quantum state after the filters will have the form\cite{*[{An outline of the derivation of the biphoton quantum state can be found in the supplementary material. For a more detailed treatment see e.g., }] [{ }] tut_biphoton,*biphoton_paraxial} $\left| \Psi \right\rangle \propto \int \d k_{x\S} \d k_{x\I} \; \psi_{\textrm{SPDC}}(k_{x\S},k_{x\I}) \; \left| k_{x\S},\omega_S \right\rangle \left| k_{x\I},\omega_I \right\rangle $, with $ \psi_{\mathrm{SPDC}}(k_{x\S},k_{x\I}) = \phi_\P(k_{x\S}+k_{x\I}) \;
 \textrm{sinc} \left(  \Delta k_z \, l_z/2 \right)$.
Here, $\Delta k_z = k_{z\P}-k_{z\S}-k_{z\I}$, $l_z$ is the thickness of the crystal, and $\left| k_{x},\omega \right\rangle$ is the single-photon state defined by the transverse component of the wave-vector $k_{x}$ and frequency $\omega$.

The JSP of the biphoton state at the two detectors in Fig.~\ref{fig:setup} is \cite{salehfourier}  
\begin{eqnarray}
    \textrm{JSP}(x_\S,x_\I) \propto \; \Big\vert \mathfrak{F}^{-1} \Big[ && h_\I (k_{x\I}) h_{2\S} (k_{x\S}) \big\{ t_o(k_{x\S}) * \big[ h_{1\S} (k_{x\S}) \nonumber \\ & &\times \psi_{\mathrm{SPDC}}(k_{x\S},k_{x\I}) \big] \big\} \Big] \Big\vert ^2 ,
    \label{eq:full}
\end{eqnarray}
where $\mathfrak{F}^{-1}$ is a two-dimensional inverse Fourier transform, $(k_{x\S},k_{x\I}) \rightarrow (x_\S,x_\I)$, and
$h$ are free space transfer functions with $h_\I (k_{x\I}) = \exp{(ik_{z\I} \;z_\I)}$, $h_{1\S} (k_{x\S}) = \exp{(ik_{z\S} \;d)}$ and $h_{2\S} (k_{x\S}) = \exp{[ik_{z\S}(z_\S-d)]}$. Finally, $*$ denotes the convolution only in $k_{x\S}$ as the object with transmission $T_o(x_\S)$ is in the signal arm. The ghost pattern $G(x_\I)$ for a bucket detector that collects all signal photons is derived from the JSP as $G(x_\I) \propto \int \d x_\S \; \textrm{JSP}(x_\S,x_\I)$. We assume a pump beam with Gaussian spatial spectrum, whose waist is located at the center of the nonlinear crystal at the plane $z=0$, where $\phi_\P \propto \exp{[-\sigma_\P^2 (k_{x\S}+k_{x\I})^2/2]}$ has a flat wave front and a width $\sigma_\P$ in position space (see zoomed out region in Fig.~{\ref{fig:setup})}.

\begin{figure}[tb]
\centering
\includegraphics[scale=1]{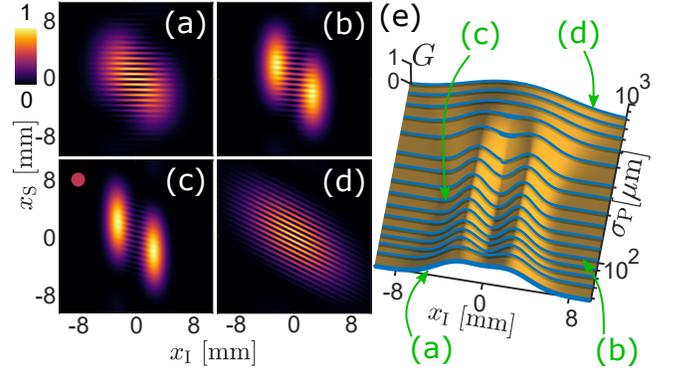}
\caption{\label{fig:numerical} (a-d) JSP$(x_\S, x_\I)$  and corresponding (e) quantum ghost pattern $G(x_\I)$ of a double-slit,  $940 \: \mu \textrm{m}$ slit separation and $50 \: \mu \textrm{m}$ slit width, located at $d = 30 \; \mathrm{cm}$ produced by a pump width $\sigma_\P$ of (a) $58 \; \mu\textrm{m}$, (b) $102 \; \mu\textrm{m}$, (c) $167 \; \mu\textrm{m}$, and (d) $800 \; \mu\textrm{m}$.}
\end{figure}

As shown in pinhole ghost imaging with a thermal source\cite{thermalpinhole}, the size of the photon source has a similar role as the pinhole size in classical optics, determining the optimal regime of imaging. This suggests that in the quantum regime, the same role can exist for the size of the biphoton source, which depends on the width of the pump beam.
To examine this premise, we numerically calculate quantum ghost imaging in a setup with a nonlinear crystal with thickness $l_z = 3 \: \textrm{mm}$, a pump with wavelength of $\lambda_\P=350 \: \textrm{nm}$, degenerate down-converted photons with $\lambda_\S=\lambda_\I=700 \: \textrm{nm}$, and detectors located at $z_\S=1.2 \: \textrm{m}$ and $z_\I=1.5 \: \textrm{m}$ from the crystal. As object, we consider a double-slit with $940 \: \mu \textrm{m}$ slit separation, with unity transmission in each slit of $50 \: \mu \textrm{m}$ width and no transmission elsewhere. In Fig.~\ref{fig:numerical}(a-d) we present the ensuing normalized JSPs calculated using Eq.~(\ref{eq:full}) 
for different sizes of the pump beam. The double-slit always results in two spots, whose separation is approximately five times larger than the slit separation. Depending on the width of the pump beam, they change their widths and begin to overlap and interfere. The interference is minimal in Fig.~\ref{fig:numerical}(c) with pump width $\sigma_\P = 167 \; \mu \textrm{m}$. In Fig.~\ref{fig:numerical}(e), we show the corresponding ghost patterns, where the cases of (a-d) are marked.  We see, that for a specific range of pump widths, two separate maxima are visible, corresponding to an image of the double-slit. We note, that the well-known quantum ghost diffraction pattern\cite{shih1} of the object could be recovered using a large pump width, as in Fig.~\ref{fig:numerical}(d), and replacing the bucket with a point detector which would measure only a horizontal cut through the JSP.

This numerical example portrays the core idea of this work. Other schemes\cite{lenslessphotonic,shih_concavemirror} have shown that a pump wave with a curved wave front, obtained by means of a lens placed before the nonlinear crystal or using a photonic crystal, can also be used for quantum ghost imaging without lenses in the paths of the biphoton. Fig.~{\ref{fig:numerical}}(c) now shows, that lensless quantum ghost imaging can be also achieved by simply using a collimated pump beam with an optimal width. This is easily seen, as the Rayleigh length{\cite{born_wolf}} of the pump, $2\pi \sigma_\P^2/\lambda_\P=50 \; \mathrm{cm}$, is much larger than the crystal thickness, $l_z = 3 \; \mathrm{mm}$.

\begin{figure}[t]
\centering
\includegraphics[scale=1]{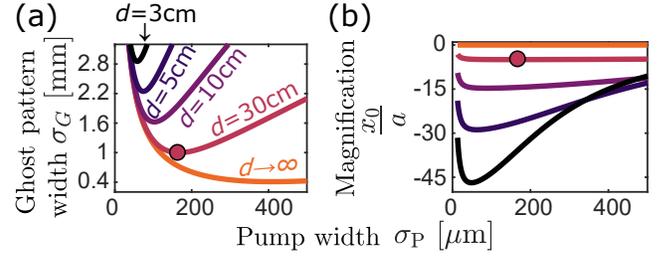}
\caption{\label{fig:quantumwidth} (a) Width $\sigma_G$ and (b) magnification of the position $x_0/a$ of the ghost pattern of a infinitesimal slit at $x_\S=a$ with respect to the pump width $\sigma_\P$ at various distances $d$ of the slit to the crystal. The circle marks the parameters of the numerical example in Fig.~\ref{fig:numerical}(c).}
\end{figure}

To find the optimal conditions for this imaging scheme, we derive a simplified analytical model. Here, we consider only one of the slits and assume it has infinitesimal width and is located at $x_\S = a$, which means that $T_o(x_\S) = \delta(x_\S - a)$. This object is put into Eq.~(\ref{eq:full}) and in paraxial approximation an analytical solution can be calculated (see supplementary material), where we approximate the sinc function appearing due to phasematching in the nonlinear crystal by a Gaussian\cite{migration}. We find a Gaussian ghost intensity pattern $G(x_\I) \propto \exp{[-(x_\I-x_0)^2/(2\sigma_G^2 )]}$ with a width $\sigma_G$ and a maximum located at $x_0$, given by
\begin{gather}
    \sigma_G = \left[ 2 \textrm{Re}\left( \alpha_1 ^{-1} \right) \right]^{-1/2},
    \label{eq:APwidth}
\\
    x_0 = a \; \textrm{Re}\left( \alpha_1 ^{-1} \alpha_2 \right) \left[\textrm{Re}\left( \alpha_1 ^{-1} \right) \right]^{-1},
    \label{eq:APmagn}
\end{gather}
where
\begin{align}
\alpha_1 &= \sigma_\P^2 + \frac{\gamma \, l_z}{\pi} \left( \lambda_\I - \lambda_\P \right) + i\frac{\lambda_\I z_\I}{2\pi} -  \left( \sigma_\P^2 - \frac{\gamma \, l_z \lambda_\P}{\pi}   \right) \alpha_2, \\ 
\alpha_2 &= \left( \sigma_\P^2 -   \frac{\gamma \, l_z \lambda_\P}{\pi} \right) \left[ \sigma_\P^2 +\frac{\gamma \, l_z}{\pi}  \left( \lambda_\S - \lambda_\P \right) + i\frac{d\lambda_\S}{2\pi} \right]^{-1}
\end{align}
with $\gamma = 0.455/4$, a constant that comes from the sinc to Gaussian approximation. Due to the bucket detector that collects all signal photons behind the object, the equations do not depend on $z_\S$, the distance of the object to the bucket detector; however, the model does depend on the location of the resolving detector $z_\I$. These distances remain at $z_\S=1.2 \: \textrm{m}$ and $z_\I=1.5 \: \textrm{m}$ throughout the manuscript. Fig.~\ref{fig:quantumwidth} shows the width $\sigma_G$ of the ghost pattern and the normalized position $x_0/a$ with respect to the width of the pump $\sigma_\P$ and the distance of the object to the crystal $d$. We observe in Fig.~\ref{fig:quantumwidth} that, for each object position $d$, the width of the ghost pattern $\sigma_G$ has a minimum at a certain pump width $\sigma_\P$. This confirms the observations of the numerical example in Fig.~\ref{fig:numerical} that uses $d=30 \: \textrm{cm}$, where the optimal case for imaging, marked with a dot in Fig.~{\ref{fig:numerical}(c)}, leads to the narrowest ghost pattern for each of the slits. In the rest of the cases, the pattern of each slit is too wide, resulting in considerable overlap between them and hence the loss of visibility of the ghost pattern. If a second infinitesimal slit would by at $x_\S=-a$, the distance between the two maxima in the ghost patterns would be $2x_0$, therefore, $x_0/a$ represents the magnification. Fig.~\ref{fig:quantumwidth}(b) verifies that for the numerical example in Fig.~\ref{fig:numerical} the magnification is approximately equal to five. It also tells that the magnification is always negative, implying that the ghost image is inverted.

Fig.~\ref{fig:quantumwidth}(a) shows, that the minimum width of the ghost pattern becomes smaller as the object is placed farther from the crystal. This value, however, does not reach zero, the behavior of $\sigma_G$ converges to approximately the orange curve in the limit where the object is very distant from the crystal, $d \to \infty$. Here, Eq.~(\ref{eq:APwidth}) can be reduced to
$ \sigma_G^2 = \sigma_0^2 + \sigma_0^{-2} \left[  z_\I \lambda_\I /(4\pi)\right]^2$.
The width of the ghost pattern of an infinitesimal slit object with the spatially resolving detector placed right after the crystal, $\sigma_0=\sigma_G(z_\I=0)$, is
\begin{equation}
    \sigma_0 = \left[\frac{1}{2} \sigma_\P^2 + \gamma \left( \frac{\lambda_\I}{\lambda_\S} \right) \left( \frac{\lambda_\P l_z}{2\pi} \right) \right]^{1/2},
    \label{eq:pinholeghost}
\end{equation} 
an expression that depends only on the parameters of the biphoton source. For a fixed value of $z_\I \lambda_\I$, the ghost pattern width $\sigma_G$ has a minimum value, namely $\sigma_G^\textrm{min}= \sqrt{2}\sigma_0$,
when $ \sigma_0^2 = z_\I \lambda_\I/(4\pi) $.
Remarkably, this result is equivalent to the optimal pinhole size $\sigma_{\textrm{pinhole}}$ in a classical pinhole camera that creates the smallest point image of a slit upon spatially incoherent illumination\cite{pinhole},  $\sigma_{\textrm{pinhole}}^2 \propto \lambda z$. Hence, $\sigma_0$ can be considered the pinhole size of quantum ghost imaging. It does not only depend on the width of the pump but also on the thickness of the crystal and the biphoton wavelengths. However, the deviation of the equivalent pinhole size $\sigma_0$ from the pump width $\sigma_\P$ due to the biphoton wavelength is small as for a pump with negligible diffraction inside the crystal $\sigma_\P^2 \gg  \lambda_\P l_z/(2 \pi)$.

Next, we analyze the magnification to complement the analogy. The magnification of the classical pinhole camera is given by geometric optics as $-z/d$ with $z$ the distance of the detector to the pinhole and $d$ the distance of the object to the pinhole. A similar relation can be found from the analytical model of the proposed ghost imaging scheme using Eq.~(\ref{eq:APmagn}), under the conditions that $\sigma_\P^2 \gg (\gamma \, l_z /\pi)(\mathrm{max}(\lambda_S,\lambda_I)-\lambda_\P)$ and $\{ d \lambda_\S /(2\pi), \; z_\I \lambda_\I / (2\pi) \} \gg \sigma_\P^2$.
The first condition ensures, that signal, idler and pump waves have negligible diffraction inside the nonlinear crystal. In this case, $\sigma_\P$ defines the size of the generated signal and idler beams inside the crystal, and hence their Rayleigh lengths are also much larger than the crystal thickness $l_z$.
The second condition states that both the object and the resolving detector are in the far-field of the biphoton source. Under these conditions, following the steps detailed in the supplementary material, we find the magnification to be $x_0/a \approx -\left(z_\I/d\right)\left(\lambda_\I/\lambda_\S \right)$. This is similar to the geometrical magnification of the classical pinhole camera but including again the biphoton wavelengths. From this expression, the magnification of the numerical example in Fig.~\ref{fig:numerical} can be quickly found $-(1.5\mathrm{m}/30\mathrm{cm})(700\mathrm{nm}/700\mathrm{nm})=-5$, see the supplementary material for an example with non-degenerate wavelengths. Noteworthy, our imaging scheme is not limited to the far-field domain. We only take the approximations to find the simple expression for the magnification to build the analogy with the classical pinhole camera.

\begin{figure}[tb]
\centering
\includegraphics[scale=1]{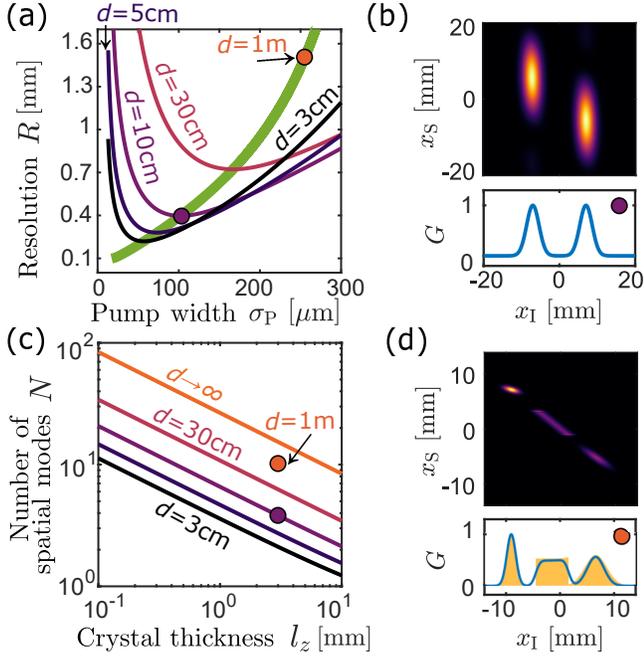}
\caption{\label{fig:res} (a) Resolution $R$ with respect to the pump width $\sigma_\P$. The green line connects the minima of curves of a wider range of object distances $d$. (c) Number of spatial modes $N$ with respect to the crystal thickness $l_z$ at various $d$.  Numerically, JSP (top) and $G$ (bottom) of a (b) double-slit and a (d) complex object (magnified transmission in yellow). Circles in (a) and (c) mark the parameters of (b) and (d).}  
\end{figure}

We further harness the analytical model of Eqs.~(\ref{eq:APwidth}) and (\ref{eq:APmagn}) to derive the transverse resolution of pinhole quantum imaging. In the following, we use Rayleigh's resolution criterion that is defined as the minimum distance between two point-like objects to distinguish them from one another\cite{born_wolf}.
A smaller ghost pattern width $\sigma_G$ of a point-like object results in a better distinction between neighboring objects, as was demonstrated in Fig.~\ref{fig:numerical}. At the same time, a larger magnification $|x_0/a|$ would results in a better distinction between two neighboring objects, as their images are further apart.
This suggests that, to optimally tell two objects apart, not only the ghost pattern width has to be taken into account but also the magnification. To test this hypothesis, we begin by taking the derived Gaussian ghost pattern from an infinitesimal slit $A$ at $x_\S=a$. For the sake of simplicity we normalize its maximum to one, resulting in the ghost pattern $G_A=\exp{[-(x_\I-x_0)^2/(2\sigma_G^2)]}$, where $\sigma_G$ is given by Eq.~(\ref{eq:APwidth}) and $x_0$ by Eq.~(\ref{eq:APmagn}). If we include another similar slit $B$ but at $x_\S=-a$, the resulting ghost pattern is symmetric with respect to $x_\I=0$ and is given approximately by $G \approx G_A+G_B$, assuming negligible interference. The two slits can be distinguished in this pattern when its visibility is above a certain threshold, here heuristically chosen to be $0.4$. That means, the intensity at $x_\I=0$ between the two maxima should be smaller than $0.4$ of the maximal intensity at $x_\I=\pm x_0$, which gives the threshold condition $G(x_\I=0)_\textrm{th}\approx G_A(0)+G_B(0)=2G_A(0)=0.4$.
Using this, the expression for $G_A$, and Eqs.~(\ref{eq:APwidth}) and (\ref{eq:APmagn}), we find the transverse spatial resolution $R$ corresponding to the minimum resolvable distance $2a$ between two identical infinitesimal slits to be
\begin{equation}
 R = 2\left\{-\ln \left[G(0)_\textrm{th}/2\right] \textrm{Re} \left( \alpha_1^{-1} \right) \right\}^{1/2}\Big\vert \textrm{Re} \left( \alpha_1^{-1} \alpha_2 \right) \Big\vert^{-1}.
\end{equation}
Fig.~\ref{fig:res}(a) displays its dependence on the pump width $\sigma_\P$ at different object distances $d$. The thick green line connects the minima of several curves over a wider range of $d$ to portray the tendency. One could naively expect from the width of the ghost pattern $\sigma_G$ in Fig.~\ref{fig:quantumwidth}(a) that the resolution $R$ could improve as the object is farther from the crystal since the minimum width becomes smaller. However, Fig.~\ref{fig:res}(a) tells the opposite, as $R$ is enhanced by the large magnification at small object distances $d$.
This is confirmed numerically by considering again the case depicted in Fig.~\ref{fig:numerical}(b), that uses a pump width of $102 \: \mu \textrm{m}$ and $d=30 \: \textrm{cm}$, and changing the position of the object to $d = 10 \: \textrm{cm}$. The resulting JSP and ghost pattern $G$ are shown in Fig.~\ref{fig:res}(b). Compared to Fig.~\ref{fig:numerical}(b), the visibility is increased due to the larger magnification. This originates from the fact that, for a fixed object distance $d$, the minimum ghost pattern width $\sigma_G$ does not coincide with the largest magnification $x_0/a$. Hence, the optimal pump width $\sigma_\P$ that results in the best resolution is not necessarily when $\sigma_G$ is minimized. This is only true for larger values of $d$ where $x_0/a$ is almost independent of $\sigma_\P$, as in the numerical example of Fig.~\ref{fig:numerical}. 

Noteworthy, the green tendency line in Fig.~\ref{fig:res}(a) hints that the closer the object is to the crystal and the smaller the pump width, the better the resolution $R$. However, the smaller the pump width, the broader its spatial spectrum becomes, in such case the used paraxial approximation does not hold. A non-paraxial formulation is a matter of future research. Additionally, $R$ will depend linearly on the object position $d$ for large values of $d$. This is because in such case, the ghost pattern width $\sigma_G$ is nearly independent of $d$, see Fig.~\ref{fig:quantumwidth}(a), and the magnification is $|x_0/a| \propto 1/d$, as already explained.

In addition to the resolution, it is also of great interest to describe the extent of the signal illumination onto the object, which limits the object size that can be imaged. This is finite and can be found from a projection of the JSP at the object position onto the signal axis (see the analytical expression in the supplementary material). This illumination size $\sigma_\S$ increases with the position of the object $d$ and decreases with the crystal thickness $l_z$. Importantly, the ratio between the illumination size and the resolution tells the maximum number of identical infinitesimal slits that can be resolved inside the illuminated region of the object, $N \equiv \sigma_\S/R$, i.e. describes the number of independent spatial modes in the object illumination \cite{spatialmodes}. Its dependence on the crystal thickness $l_z$ is displayed in Fig.~\ref{fig:res}(c) for various $d$, each at a pump width $\sigma_\P$ that minimizes the resolution as described by the green line in Fig.~\ref{fig:res}(a). To increase the number of spatial modes $N$, a thinner crystal can be used, similar to other quantum imaging schemes\cite{shih2,undetectedPhotons}, as it allows a larger range of transverse wave-vectors\cite{ultrathincrystal}. Also, the object could be put farther away from the crystal. Here, the increase of $N$ with $d$ has a limit due to the linear dependence of both $\sigma_\S$ and $R$ on $d$ for very distant objects. Finally, we found that photon-pairs with non-degenerate wavelengths show a minor improvement in the resolution and number of modes, see the supplementary material for some examples.

Lastly, we sum up the main features of the proposed setup with a ghost image of a more complicated object with various shapes and transmissions, see Fig.~\ref{fig:res}(d). This object is placed at $d= 1 \; \mathrm{m}$ and we use a pump width $\sigma_\P = 258 \; \mu \mathrm{m}$ that optimizes the ghost image resolution to $R = 1.5 \; \textrm{mm}$, allows $N=10$ spatial modes and has a magnification $x_0/a = -1.2$. Moreover, unlike the setup with a pseudo-thermal source{\cite{thermalpinhole}}, we see in the JSP of this example that an integrating bucket detector is necessary to show the whole illuminated section of the object in the ghost pattern, a point detector would not be able to detect signals from all objects as it would just ``take a horizontal thin slice'' of the JSP.

To conclude, we proposed a quantum ghost imaging scheme without lenses in the biphoton arms by means of a collimated pump beam with an optimal size. This imaging scheme is best suited for applications where lenses for the  biphoton wavelengths are less available and a high transverse resolution is not required. We demonstrated that the proposed scheme is analogous to the classical pinhole camera where the biphoton source plays the role of the pinhole and derived its spatial resolution and number of spatial modes.


See the supplementary material for further details of the analytical model and examples with non-degenerate photons.

\begin{acknowledgments}
We thank E. Santos and V. Gili for their insightful comments. This work was supported by the Thuringian Ministry for Economy, Science, and Digital Society; the European Social Funds and the European Funds for Regional Development (2017 FGR 0067, 2017 IZN 0012); the German Federal Ministry of Education and Research (FKZ 13N14877, FKZ 03ZZ0434) and the Deutsche Forschungsgemeinschaft (DFG, German Research Foundation, project ID 407070005).
\end{acknowledgments}

\section*{Data availability}
The data that supports the findings of this study are available within the article and its
supplementary material.

This article may be downloaded for personal use only. Any other use requires prior permission of the author and AIP Publishing. This article appeared in A. Vega \emph{et al.,} Appl. Phys. Lett. \textbf{117}, 094003 (2020) and may be found at https://doi.org/10.1063/5.0012477


\bibliography{aipsamp}

\end{document}